\documentclass[reprint,superscriptaddress, amsmath,amssymb, aps,prl,floatfix,]{revtex4-1}
\usepackage[utf8]{inputenc}
\usepackage{color}
\usepackage{graphicx}% Include figure files
\usepackage{placeins}
\usepackage[caption=false]{subfig}
\usepackage{xspace}
\usepackage{xfrac}
\usepackage{bm}% bold math
\usepackage[hidelinks]{hyperref}% add hypertext capabilities
\newcommand{\ymo}{$h$-YMnO$_3$\xspace}

\begin{document}
\title{Classical spin liquid or extended critical range in \texorpdfstring{\textit{h}-YMnO$_3$}{h-YMnO3}?}
\author{Sofie Janas}
\affiliation{Nanoscience Center, Niels Bohr Institute, University of Copenhagen, 2100 Copenhagen \O, Denmark}
\affiliation{Institute of Physics, École Polytechnique Fédérale de Lausanne, CH-1015 Lausanne, Switzerland}
\author{Jakob Lass}
\affiliation{Nanoscience Center, Niels Bohr Institute, University of Copenhagen, 2100 Copenhagen \O, Denmark}
\affiliation{Laboratory for Neutron Scattering, Paul Scherrer Institute, 5232 Villigen, Switzerland}
\author{Ana-Elena \textcommabelow{T}u\textcommabelow{t}ueanu}
\affiliation{Nanoscience Center, Niels Bohr Institute, University of Copenhagen, 2100 Copenhagen \O, Denmark}
\affiliation{Institut Laue-Langevin, Grenoble Cedex 9 38042, France}
\author{Morten L. Haubro}
\affiliation{Nanoscience Center, Niels Bohr Institute, University of Copenhagen, 2100 Copenhagen \O, Denmark}
\author{Christof Niedermayer}
\affiliation{Laboratory for Neutron Scattering, Paul Scherrer Institute, 5232 Villigen, Switzerland}
\author{Uwe Stuhr}
\affiliation{Laboratory for Neutron Scattering, Paul Scherrer Institute, 5232 Villigen, Switzerland}
\author{Guangyong Xu}
\affiliation{NIST Center for Neutron Research, National Institute of Standards and Technology, Gaithersburg, Maryland 20899, USA}
\author{Dharmalingam Prabhakaran}
\affiliation{Clarendon Laboratory, Department of Physics, University of Oxford, Oxford OX1 3PU, United Kingdom}
\author{Pascale P. Deen}
\affiliation{European Spallation Source ESS ERIC, Box 176, SE-221 00 Lund, Sweden}
\affiliation{Nanoscience Center, Niels Bohr Institute, University of Copenhagen, 2100 Copenhagen \O, Denmark}
\author{Sonja Holm-Dahlin}
\affiliation{Nanoscience Center, Niels Bohr Institute, University of Copenhagen, 2100 Copenhagen \O, Denmark}
\affiliation{ISIS Facility Rutherford Appleton Laboratory Chilton, Didcot, OX 11 OQX, United Kingdom}
\author{Kim Lefmann}
\affiliation{Nanoscience Center, Niels Bohr Institute, University of Copenhagen, 2100 Copenhagen \O, Denmark}
\date{\today}

\begin{abstract}
Inelastic neutron experiments on the classical triangular-lattice geometrically frustrated antiferromagnet \ymo reveal diffuse, gapless magnetic excitations present both below and far above the ordering temperature, $T_{\rm N}$.
The correlation length of the excitations increases as the temperature approaches zero, bearing strong resemblance to critical scattering. We model the scattering as critical spin-spin correlations in a two-dimensional magnetic ground state, and we speculate that this may provide a general framework to understand features typically attributed to classical spin liquids.
\end{abstract}
\maketitle

Frustrated magnetism, in which competing exchange interactions suppresses magnetic order, is ubiquitous within condensed matter physics and one of the most heavily discussed topics is quantum spin liquids (QSL). Theoretically, QSLs are defined as fluid-like states, where the spins are highly correlated and continue to fluctuate down to temperatures of absolute zero \cite{Savary2017,Moessner2018,Wen2002,Balents2010}, although experimental verification is still contentious. Many frustrated compounds show features reminiscent of QSLs, but order magnetically at finite temperatures, which excludes them as QSL candidates. These compounds are referred to as classical spin liquids \cite{Balents2010}, cooperative paramagnets \cite{Moessner2018}, or merely having spin-liquid-like phases \cite{Park2003}, although no clear definition of these terms is readily available. Here, we show that such spin-liquid-like features in the classical triangular lattice Heisenberg antiferromagnet with nearest-neighbour exchange interactions (TLHA) of hexagonal YMnO$_3$ can be modelled as arising from critical scattering in a vastly extended temperature regime due to its geometrical frustration. This may provide a general framework for the understanding of frustration in TLHA and similar systems.

\begin{figure*}[htb]
\includegraphics{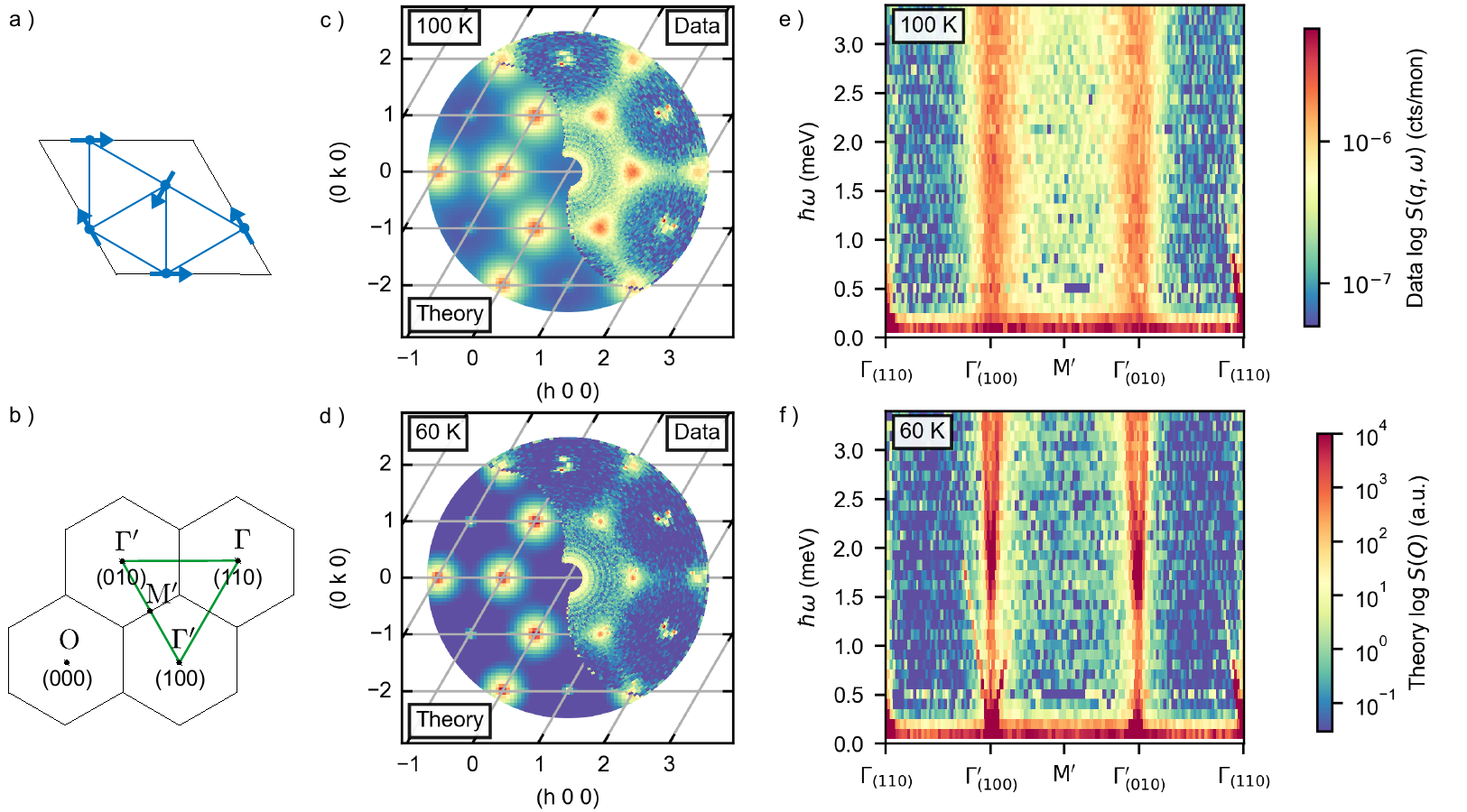} 
\caption{(a) The spin structure of \ymo in the two-dimensional triangular lattice with Mn spins in the 120$^\circ$ structure. (b) Sketch of reciprocal lattice in $(hk0)$-plane. $\Gamma^\prime$ is a magnetic Bragg peak, $\Gamma$ is a structural Bragg peak, and M$^\prime$ is the center of the Brillouin edge between neighboring magnetic Bragg reflections. Green lines indicate the direction of the $\mathbf{Q}$ cut used in subplots (e-f). (c-f) Color maps of inelastic neutron scattering data on \ymo measured above and below $T_N$ on a common, logarithmic intensity scale, created with the software MJOLNIR \cite{Lass2020MJOLNIR}. Subplots (c-d) show constant energy cuts at $\hbar \omega = 1 \pm 0.2$~meV for $T=100$~K and $T=60$~K compared with numerical simulations of the scattering as critical fluctuations. The model is described in the text, and plotted on its own logarithmic intensity scale. (e-f) $(\mathbf{Q},\hbar\omega)$ cuts of the same experimental data along the green line in subplot (b).} 
    \label{fig:cameamaps}
\end{figure*}

The triangular lattice is one of the simplest geometrical motifs to exhibit frustration, although the TLHA has been shown to have an ordered ground state at $T=0$ in both the quantum and classical spin case \cite{Bernu1992,Capriotti1999}. In this paper, we study the hexagonal rare-earth manganite YMnO$_3$, which has stacked triangular lattice planes and is a close approximation to a TLHA with classical spins. \ymo is also widely studied due to its type-I multiferroic nature. In \ymo,  $S=2$ Mn$^{3+}$--ions form a two-dimensional triangular lattice, see Fig.~\ref{fig:cameamaps}(a), separated by interlayer Y and O ions. \ymo has a Curie-Weiss temperature of $\theta_{CW}=-545$~K \cite{Park2003}, and a magnetic phase transition to a three-dimensionally ordered antiferromagnetic state at $T_N=71$~K \cite{Roessli2005,Holm-Dahlin2018}. Its magnetic interactions are dominantly an antiferromagnetic Heisenberg nearest-neighbor exchange, $J=2.4$~meV, then a weaker easy-plane anisotropy within the hexagonal plane with $D/J\approx0.13$, and finally even weaker interlayer interactions, $J_z/J\approx0.06$ \cite{Holm2018}, thus making \ymo highly two-dimensional and TLHA-like. The magnetic ordering below $T_N$ within the triangular planes is the $120^\circ$ spin structure shown in Fig.~\ref{fig:cameamaps}(a).

Diffuse scattering in \ymo observed via inelastic neutron scattering, were previously reported by a number groups on both powders and single crystals \cite{Sato2003,Chatterji2007,Kozlenko08, Demmel2007, Lonkai2003, Roessli2005}. The diffuse scattering was ascribed to a spin liquid state. 
However, no attempt at a coherent understanding and modeling of these features has been made, and even temperature and energy dependencies of the signal remains little explored.

We report an inelastic neutron scattering study on \ymo, thus directly probing the magnetic dynamic structure factor, $S(\mathbf{Q},\omega)$ as a function of scattering wavevector $\mathbf{Q}$ and energy transfer $\hbar \omega$ \cite{Shirane2013}. Our experiments were performed on a 3.4~g single crystal of \ymo  aligned in the $(hk0)$ crystal plane. To obtain an overall picture of the excitations, we used the cold-neutron spectrometer CAMEA (PSI) \cite{Groitl2016,Lass2020CAMEA,Birk2014} that utilizes multiplexing analyzers and a large detector array to obtain quasi-continuous coverage in three-dimensional $(h, k, \hbar\omega)$ space.
The data, Fig.~\ref{fig:cameamaps}(c-f), is taken at $T = 60$~K ($<T_{\rm N}$) and at $T=100$~K ($>T_{\rm N}$). Constant-energy maps with $\hbar\omega=1.0 \pm 0.2$~meV are shown in Figs.~\ref{fig:cameamaps}(c-d), superimposed on a theoretical simulation described later. Figs.~\ref{fig:cameamaps}(e-f) show $(\mathbf{Q},\hbar\omega)$ cuts of the same data across two magnetic Bragg reflections. The cut direction is sketched in Fig. \ref{fig:cameamaps}(b).

In the constant-energy $(hk0)$-map for 100~K in Fig.~\ref{fig:cameamaps}(c) we observe a diffuse signal of hexagonal symmetry. Broad peaks of higher intensity reside at $\Gamma^\prime$, consistent with the location of magnetic Bragg peaks below $T_{\rm N}$.
These peaks are connected with sheets of diffuse scattering crossing the $M^\prime$ points to form a hexagonal pattern. The signal is strongest for small $q$, as expected due to the magnetic form factor \cite{Shirane2013}. Figure \ref{fig:cameamaps}(d) shows that the scattering for $T=60$~K ($< T_{\rm N}$) is very similar to the scattering at 100~K, although less intense, and the peaks are more narrow, indicating longer ranged correlations. In both data sets so-called Currat-Axe spurions  \cite{Shirane2013,Supplementary} are present. Figs.~\ref{fig:cameamaps}(c-d) show them as two narrow, bright peaks, close to the Bragg peaks, and in Figs.~\ref{fig:cameamaps}(e-f) they are seen as sharp, linearly-dispersing streaks.

In the $(\mathbf{Q},\hbar\omega)$ cut at 100~K, Fig.\ \ref{fig:cameamaps}(e), the $\Gamma^\prime$ excitation is seen as broad rods with approximately constant intensity and width for increasing energy. The rods are gapless within the 0.2~meV energy resolution. The M$^\prime$ excitation is seen as increased intensity between the two $\Gamma^\prime$ points, and also appears featureless with energy-independent intensity. Below $T_{\rm N}$, Fig.~\ref{fig:cameamaps}(f), a spin wave is seen as a steep parabolic shape that has its bottom at the $\Gamma^\prime$ point, at $\hbar \omega = \Delta\approx 1.7$~meV. The spin wave gap $\Delta$ is due to the single-ion anisotropy of \ymo \cite{Holm-Dahlin2018,Petit2007,Sim2016}. However, the intensity below $\Delta$ is not associated with any spin wave, and we recognize it as the same diffuse signal as above $T_{\rm N}$, although it is significantly narrowed below $T_{\rm N}$. 

Based on these maps, we carried out detailed studies of the temperature and energy dependencies of the excitations using combined data from triple-axis spectrometers (TAS). The thermal-neutron TAS Eiger (PSI) \cite{Stuhr2017} and IN3 (ILL) were used to study the signal at $M^\prime$, while the cold-neutron TAS SPINS (NIST) was used to study the excitation at $\Gamma^\prime$, using the better energy resolution to access the excitation below the spin wave gap in the ordered phase. The intensity has been cross-normalized. See also \cite{Supplementary}.

\begin{figure}[htb]
    \includegraphics{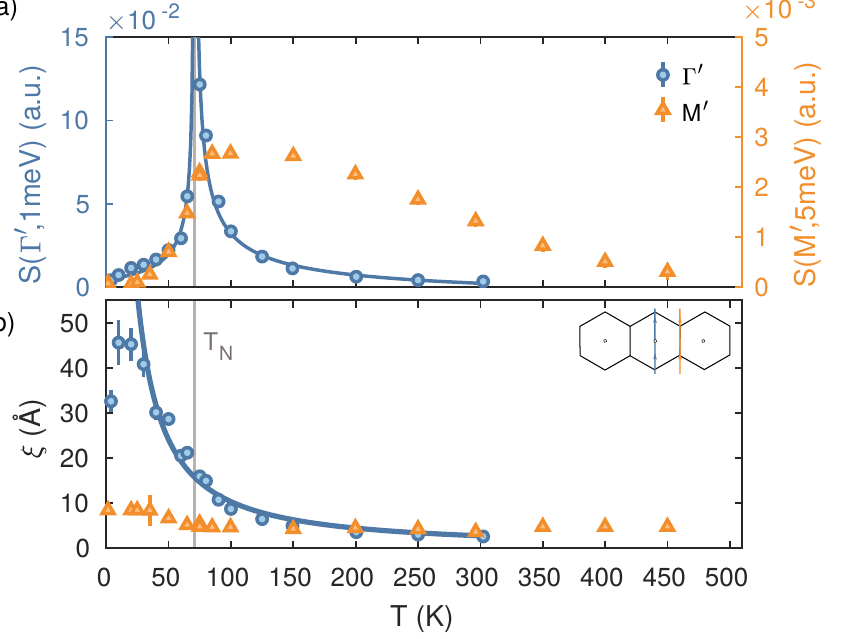}
    \caption{Temperature dependence of the inelastic neutron scattering data for (a) the integrated intensity and (b) the correlation length for $\Gamma^\prime$ and M$^\prime$. Scan directions are shown schematically in (b). Full blue lines are power-law fits to $\Gamma^\prime$ data.}
    \label{fig:QFit}
\end{figure}

We investigated the temperature dependence of the intensities and peak widths by constant-energy scans through $\Gamma^\prime$ with $\hbar \omega =1$~meV at SPINS, and through M$^\prime$ with $\hbar \omega=5$~meV at Eiger and IN3. The scan directions are shown in the insert of Fig.~\ref{fig:QFit}(b). Each data set is fitted with a Lorentzian, and the integrated intensity and real-space correlation lengths $\xi$ are shown in Fig.~\ref{fig:QFit}(a-b) for both $\Gamma^\prime$ and M$^\prime$.

The integrated intensity at $\Gamma^\prime$, Fig.~\ref{fig:QFit}(a), increases quickly when approaching $T_{\rm N}$ from either side. To illustrate the resemblance to a divergence at the phase transition, a power law is fitted to the $\Gamma^\prime$ data. The $\Gamma^\prime$ excitation is well-defined at both the highest (301~K~$\approx 4.2~T_{\rm N}$) and lowest (4~K~$\approx 0.06~T_{\rm N}$) temperatures accessed, and is more intense than the M$^\prime$ excitation.
The integrated intensity at M$^\prime$ is at background level below 25~K, and then quickly increases in intensity upon heating through $T_{\rm N}$. Above $T_{\rm N}$, the intensity decreases very slowly, reaching background level at 450~K, close to $|\theta_{CW}|=545$~K.
The size of the correlation length for the $\Gamma^\prime$ excitation above $T_N$ is $\xi\approx 4$~\AA, see Fig.\ \ref{fig:QFit}(b), similar to the inter-atomic distance between neighboring Mn atoms, 3.53~\AA. When cooling below $T_{\rm N}$, the correlation length increases smoothly to 45~\AA. Curiously, no divergence in $\xi$ is found at $T_{\rm N}$. However, neglecting the 3 lowest points, the $\Gamma^\prime$ data fits a power law diverging at $T=0$. The correlation length of the M$^\prime$ excitations (Fig.~\ref{fig:QFit}b) is approximately constant at 4~\AA~for the entire temperature range, matching the correlation length of the $\Gamma^\prime$ point excitation at higher temperatures.

\begin{figure}[htb]
    \includegraphics{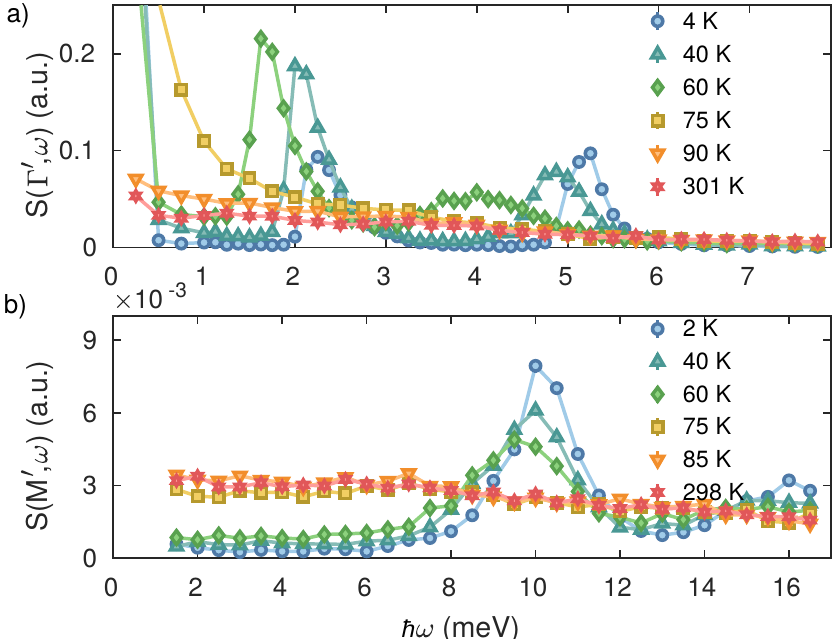}
    \caption{Energy dependence of diffuse excitations at $\Gamma^\prime$ (a) and at M$^\prime$ (b) measured with inelastic neutron scattering.}
    \label{fig:energy}
\end{figure}

The energy dependence of the diffuse excitations was investigated at different temperatures by constant-$Q$ scans at $\Gamma^\prime$ and M$^\prime$, see Fig.~\ref{fig:energy}.
The $\Gamma^\prime$ point data at 4~K, Fig.~\ref{fig:energy}(a), show two magnon modes at 2.2~meV and at 5.2~meV \cite{Oh2016,Holm2018}. With increasing temperatures the magnons soften \cite{Holm-Dahlin2018}, but in addition, we observe diffuse intensity. At 75~K, just above $T_{\rm N}$, we observe quasielastic, critical fluctuations close to the three-dimensional phase transition. This behavior subsides already for 90~K, above which the diffuse excitation has roughly constant intensity up to 4~meV, where it eventually vanishes at 6~meV. For M$^\prime$ at 2~K, see Fig. \ref{fig:energy}(b), magnons are seen at 10~meV and 16~meV \cite{Oh2016,Holm2018}. Increasing temperature broadens and weakens the magnons, and we observe an increase of the diffuse signal. Upon crossing $T_{\rm N}$ a sudden jump is observed in the intensity, after which the intensity is almost energy independent. This behavior is unchanged up to 300~K.

The inelastic signal at $\Gamma^\prime$ exhibits a divergence-like spike in intensity at $T_{\rm N}$ and a spike in correlation length at low temperatures with similarities to a divergence at $T=0$. The smooth energy dependence of the fluctuations above $T_{\rm N}$ corresponds to fluctuations at all timescales below the characteristic energy scale $zJS$.
These features are spin-liquid-like due to the resemblance to the diffuse fractionalization which is an experimental fingerprint of spin liquids \cite{Moessner2018, Savary2017}. However, the diverging intensities and correlation lengths also resemble features of critical scattering: around a phase transition there will be diffuse, quasielastic scattering whose intensity and correlation length diverge at the transition temperature \cite{Shirane2013,CollinsBook89}. For unfrustrated magnets this typically happens in a temperature range of $\sim$5--10\% of $T_N$, where diffuse scattering sharpens to form Bragg reflections. If the measured excitations follow this trend, they should have a pronounced intensity variation along the $l$ direction. 

To investigate this dimensionality of the diffuse excitations, we investigated on IN3 the M$^\prime$ excitation for different $l$ values on a smaller sample aligned in the $(hhl)$ scattering plane, see \cite{Supplementary}. Both integrated intensity and width appear independent of $l$ in the temperature range between 70 and 100~K, see Fig.~4. This indicates that the diffuse scattering is a two-dimensional phenomenon within the triangular lattice plane, and does not contribute to the formation of Bragg reflections from three-dimensional order.
\begin{figure}[htb]
    \centering
    \includegraphics{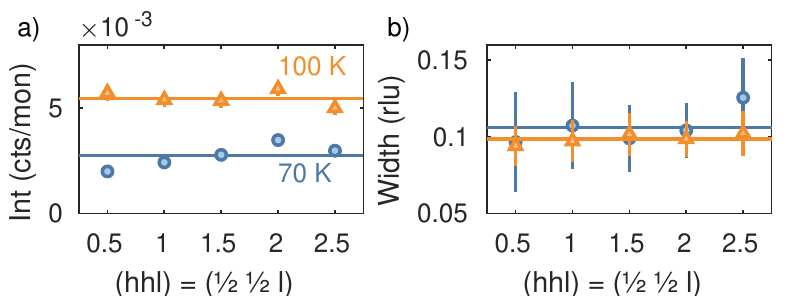}
    \caption{Scattering intensity and width of the M$^\prime$ excitation measured for different $l$-values. Scans across the excitation at 70~K and 100~K was fitted with a Lorentzian, and shown here is the (a) integrated intensity and (b) width. Full lines indicate mean value of parameters.}
\label{fig:IN3-lpar}
\end{figure}
Our findings prove that the M' signals do not arise from simple critical scattering. However, they could be caused by critical scattering from the suppressed two-dimensional order, which would appear at $T=0$ in an idealized TLHA system. In this scenario, the critical range is vastly extended (compared to the critical range around $T_N$) due to the strength of the interactions, while the order is suppressed by geometric frustration. 

Investigating this similarity, we model our data as critical scattering from correlated clusters populated with spins in the two-dimensional 120$^\circ$ ground state that oscillate coherently. We do this by calculating the static structure factor,

\begin{align}
       S(\mathbf{Q})= f(Q)^2 \sum_{l=0}^N  \sum_{\alpha,\beta}   e^{i \mathbf{Q} \cdot \mathbf{r}_{l}} \left(\delta_{\alpha, \beta}-\hat{Q}_{\alpha} \hat{Q}_{\beta}\right)
 \left\langle S_{0}^{\alpha} S_{l}^{\beta}\right\rangle, 
 \label{eq:Sq}
\end{align}
where $N$ is the number of spins, $\alpha,\beta$ refers to the $x,y,z$ vector components, $f(Q)$ is the magnetic form factor for Mn$^{3+}$, $\mathbf{r}_l$ their position in the lattice, $S_l^\alpha$ refers to their spin component, and the brackets $\langle \dots \rangle$ refers to a $4\pi$ rotational average. In this simple model, the Q-dependence of the dynamical structure factor equals that of the static structure factor.

The cluster structure is typically taken as the minimal geometrical motif in the lattice \cite{Lee2002}, which in our case is the spin trimer. However, since our correlation lengths extend beyond the nearest-neighbor, we instead simulate a larger supercell. Under the assumption of critical scattering in the two-dimensional planes, we incorporate the spin-spin correlation by the critical form $\langle S_0 S_l \rangle \propto \exp(-|\mathbf{r}_{0l}|/\xi)$ \cite{CollinsBook89}. Using the temperature-dependent correlation lengths determined for $\Gamma^\prime$, $\xi(100~\textrm{K})=8.6$~\AA~ and $\xi(60~\textrm{K})=20.5$~\AA, we set up a supercell of 40$\times$40 unit cells consisting of 4800 spins to achieve the $S(\mathbf{Q})$ spectra shown alongside the data in Fig.~\ref{fig:cameamaps}(c-d). The simulations qualitatively reproduces the experimental data very well: there are peaks at the $\Gamma^\prime$ point for both correlation lengths with a broader signal for $\xi=8.6$~\AA~as expected, and the peaks are connected by weaker bridge-like intensity. In Fig.~\ref{fig:cameamaps}, both measured and simulated $S(\mathbf{Q})$ are plotted on a logarithmic color scale. However, the simulated scale span more decades than the experimental data. We attribute this to the simplicity of our model, neglecting energy dependencies.
 Still, the model allows us to conclude that the $\Gamma^\prime$ and M$^\prime$ excitations arise from the same origin despite the different temperature and energy dependencies.
Additionally, the large cluster with exponentially decaying spin-spin correlations much better captures the experimental features than the simple trimer model \cite{Supplementary}.

We envision the following physical picture for the short-range dynamics in \ymo: Mn$^{3+}$ spins on the lattice develop local 120$^\circ$ in-plane correlations below $\theta_{\rm CW}$, giving rise to diffuse scattering.
These magnetic clusters are subject to thermal noise and fluctuate on all relevant time scales. With decreasing temperature, the typical size of the spin clusters, $\xi$, increases, causing increased scattering intensity. At $T_N$ a three-dimensional magnetic ordering occurs due to the coupling between the Mn layers, and collective spin waves emerge. However, the spin clusters co-exist with the spin waves, even as the temperature decreases. Indeed, in the limit $T \rightarrow 0$, the size of the clusters increases, while the intensity vanishes.  Since the fluctuations of the clusters happen at all time scales, in contrast to the well-defined frequencies of spin waves, we believe that these two dynamic phenomena are decoupled and at most weakly interacting. 

Scattering patterns reminiscent of our experimental data have been reported in theoretical studies of the purely two-dimensional TLHA, which were based on exponential tensor renormalization group theory for quantum spins \cite{Chen2019}, and Monte Carlo simulations on classical spins \cite{Okubo2010}. Both studies obtain diffuse scattering maps consistent with ours. Notably, \citet{Chen2019} observe a temperature dependence of the scattering intensities analogous to our experimental data, when ignoring the presence of a finite-temperature magnetic phase transition. We take these theoretical results as further evidence that the diffuse scattering stems purely from the two-dimensional 120$^\circ$ order. Some groups have put forth more exotic explanations for the cause of the diffuse scattering, such as $Z_2$ spin vortices \cite{Okubo2010,Kawamura1984} and chiral roton-like excitations \cite{Chen2019}. However, we do not find any compelling evidence for either of these as explanation in our data: chiral rotons should be gapped \cite{Chen2019}, whereas our data shows gapless excitations within the experimental resolution, and a $Z_2$ spin vortex cannot be accommodated with our short observed correlation lengths. Experimental observations of diffuse excitations similar to ours in geometrically frustrated compounds have often been referred to as spin liquids, however we show that in \ymo the excitations can be satisfactorily modeled by critical spin fluctuations existing in a vastly extended critical region due to the frustration. Thus, it seems worth contemplating whether such features in other compounds can also be modeled by critical fluctuations, which may enable a more precise definition of the term 'classical spin liquid'.

The vastly extended critical range has profound consequences on the physical properties of \ymo. This is evidenced by the anomalous thermal conductivity $\kappa$, observed by \citet{Sharma04} in the temperature range between 20~K and 300~K. These results are not explained by the usual phononic terms, whence the authors proposed the anomalies to be caused by short-ranged magnetic excitations coupled to the heat-carrying phonons via exchange-striction \cite{Sharma04}. Their observed anomalous temperature range corresponds very well to the temperature range where we observe strong critical fluctuations. For this reason, we speculate that it is indeed these short-ranged critical fluctuations that dynamically couple to the phonons and cause the thermal conductivity anomalies. 
Similarly, anomalous values for the critical exponent $\beta$ that do not correspond to any of the well-known universality classes have been reported for \ymo \cite{Holm-Dahlin2018} and other TLHA materials \cite{Collins97,Kawamura1998}. We speculate that this is caused by the presence of the additional, two-dimensional critical fluctuations in the extended critical range which moves the phase transition beyond the simple Landau regime. 

In summary, our experiments on \ymo provide evidence for two-dimensional critical scattering as the origin of low-lying, diffuse scattering observed in triangular-lattice antiferromagnets.
We show good agreement with a critical model based on exponentially decaying correlations in an extended critical range due to geometric frustration. We believe that these detailed results may contribute to establishing a more detailed understanding of features attributed to classical spin liquids. 

\begin{acknowledgments}
This work was supported by the Danish Agency for Research and Innovation through DANSCATT and the UK Engineering and Physical Sciences Research Council Grant no. EP/M020517/1. The neutron experiments were performed at the Paul Scherrer Institute (CH), at the Institute Laue Langevin (F), and at the NCNR neutron center (US). We thank Nicola Spaldin and Bruce Gaulin for helpful discussions.
\end{acknowledgments}
\bibliography{bib}

%\documentclass[reprint,superscriptaddress, amsmath,amssymb, aps,prl,floatfix,]{revtex4-1}
%\usepackage[utf8]{inputenc}
%\usepackage{color}
%\usepackage{graphicx}% Include figure files
%\usepackage[caption=false]{subfig}
%\usepackage{placeins}
%\usepackage{blindtext}
%\usepackage{xspace}
%\usepackage{bm}% bold math
%\usepackage{pstricks}
%\usepackage[hidelinks]{hyperref}% add hypertext capabilities
%\graphicspath{{figures/}}
%\newcommand{\ymo}{$h$-YMnO$_3$\xspace}

%\begin{document}
%\title{Supplementary Material}
%\maketitle
\appendix
%\FloatBarrier
\section{APPENDIX}
\section{Sample details}
At room temperature, \ymo is in the hexagonal $P6_3cm$ space group with $\alpha=\beta=90^\circ$, $\gamma=120^\circ$, and lattice parameters $a=b=6.11$~\AA~, and $c=11.44$~\AA~ \cite{Lorenz2019}.
Below the antiferromagnetic phase transition at $T_{\rm N}=71$~K, the system orders magnetically in the $120^\circ$ spin structure with the low-temperature magnetic symmetry group P6$'_3cm'$ \cite{Fiebig2000,Howard2013,Holm2018}. 

Our main experiments were performed on a 3.4~g single crystal sample of \ymo grown by the floating zone method \cite{Lancaster2007} and aligned in the $(hk0)$-plane. In addition, the study of the dimensionality was performed on a smaller 1.3~g crystal also grown by the floating zone method and aligned in the $(hhl)$-plane.  

\section{Instrumental details}
Data was taken with three different conventional triple-axis spectrometers (Eiger, PSI; SPINS, NIST; IN3, ILL), as well as the newly commissioned multiplexing triple-axis spectrometer CAMEA, PSI. Details for each experiment is given below.\\

CAMEA was configured with a wide-angle Beryllium filter covering all detector banks, where we used incoming energies of $E_i=5.25$~meV and 6.85~meV. These datasets were then stitched together. The temperature was controlled using a standard Orange cryostat. All CAMEA data was treated and visualized using the MJOLNIR software \cite{Lass2020MJOLNIR}. Data is presented in Fig. 1 in the main text. The three-dimensional $(h,k,\hbar \omega)$ data is shown as both constant-energy slices in $(h,k)$, and as energy-Q slices $(\mathbf{Q},\hbar \omega)$ along a particular direction in $\mathbf{Q}$. For the $(h,k)$ map, data with energy $\hbar \omega = 1 \pm 0.2$~meV is shown, with a polar binning of 0.03 \AA$^{-1}$ bin tolerance in both x and y directions. For the $(\mathbf{Q},\hbar \omega)$ slice, data is presented with a binning of 0.1~meV in energy transfer, and 0.05~Å$^{-1}$ in $q$ perpendicular to the direction shown.

The CAMEA data contains spurions emerging from Bragg reflections, and dispersing linearly in energy. These are known as Currat-Axe spurions and stem from accidental elastic scattering, as described in great detail in \citet{Shirane2013}. In short, whenever two out of the three crystals involved (monochromator, sample, and analyzer) are set to Bragg reflect, then a weak scattered signal from the third crystal (such as incoherent or thermal diffuse scattering) may be detected, and though it is an elastic signal it shows up at a spurious inelastic energy transfer. The two spurions in the data are from accidental scattering in the monochromator and analyzer, while accidental scattering in the sample belongs to the elastic line.\\

Detailed temperature- and energy scans were taken using three different triple axis spectrometers (TAS) using both thermal and cold neutrons. This allowed us to access different energy ranges with resolution functions of different sizes. Since the diffuse signals are so broad in $q$ (and the same sample was used in all experiments in the $(h,k)$-plane), it is possible to cross-normalize the observed intensities from different experiments using a standard scan.

\begin{figure*}[htb]
    \centering
    \includegraphics{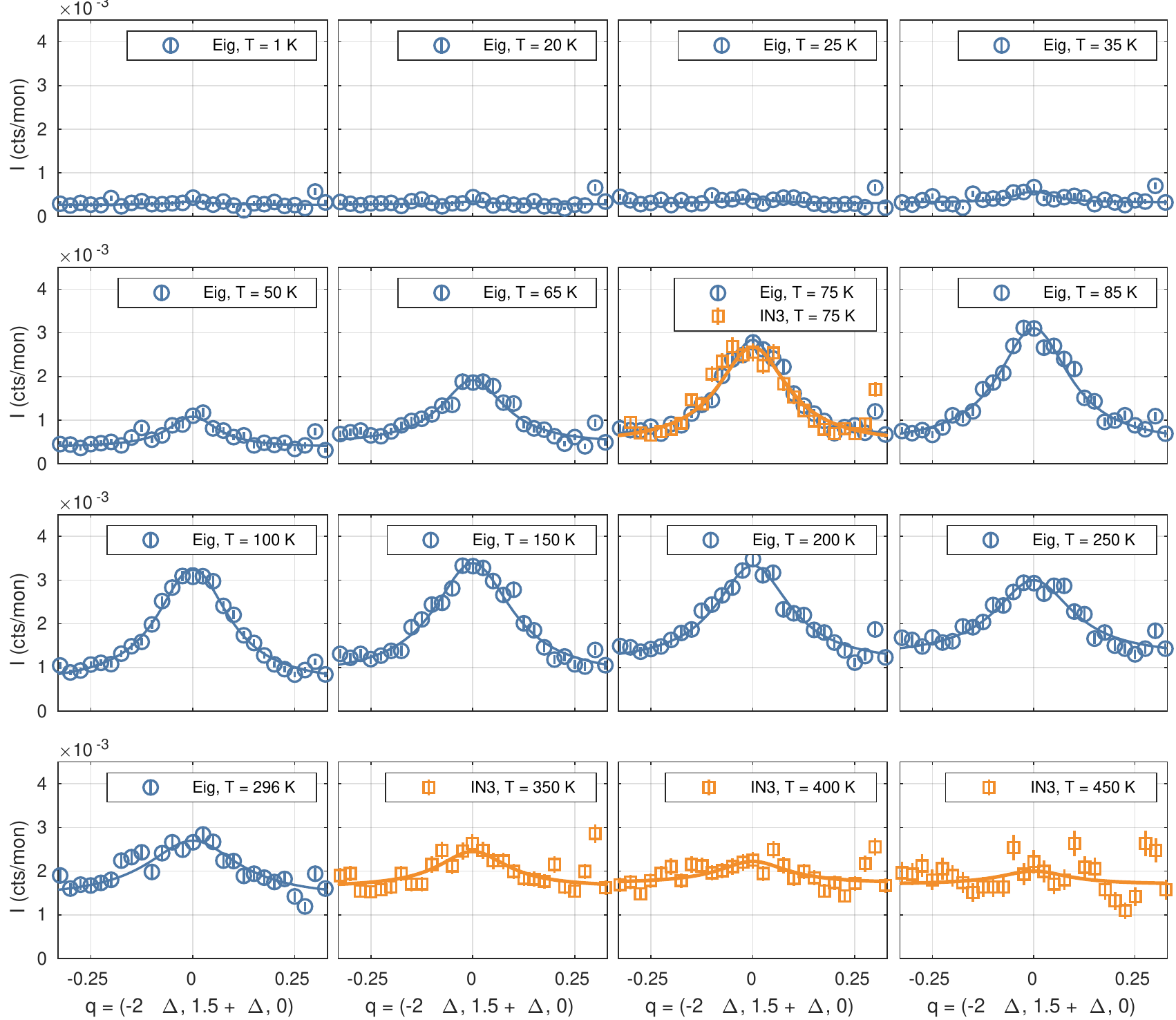}
    \caption{Raw scans for the M$^\prime$ point excitation at $\hbar \omega = 5$~meV taken with Eiger (blue) and IN3 (orange) for various temperatures, along with Lorentzian fits. IN3 scans are shown cross-normalized.}
    \label{fig:Mscans}
\end{figure*}
The excitations around the M$^\prime$ point was investigated using the thermal neutron TAS Eiger \cite{Stuhr2017} (PSI). We utilized fixed final energy of $E_{\rm f}=14.7$~meV, an orange cryostat, and a PG filter before a horizontally focusing analyzer.
Data from this experiment was combined with data from an experiment on the thermal neutron TAS IN3 (ILL) using a cryofurnace to access higher temperatures. IN3 was also configured with fixed energy of $E_{\rm f}=14.7$~meV and a PG filter before a horizontally focusing analyzer.
The cross normalization was performed between two scans across the M$^\prime$ point at $(h,k,l)=(0,1.5,0)$ at $T=75$~K. After fitting a Lorentzian to each dataset, a normalization factor for the IN3 intensity was obtained so that it matched the Eiger intensity for this scan. This intensity normalization factor was then used for all IN3 scans, the results of which is shown in Fig. 2 in the main article. The width of the diffuse excitations at the M$^\prime$ points are 0.1--0.3~\AA$^{-1}$, while the resolution of thermal TAS instruments is roughly 0.03~\AA$^{-1}$. For the $\Gamma^{\prime}$ point the width varies between 0.02--0.4~\AA$^{-1}$, and the resolution of the cold TAS instruments is rouhgly 0.003~\AA$^{-1}$. Thus the signals are in no way resolution-limited, and the simple cross-normalization can then be performed. The raw scans and their Lorentzian fits can be seen in Fig. \ref{fig:Mscans}.

\begin{figure*}[htb]
    \centering
    \includegraphics{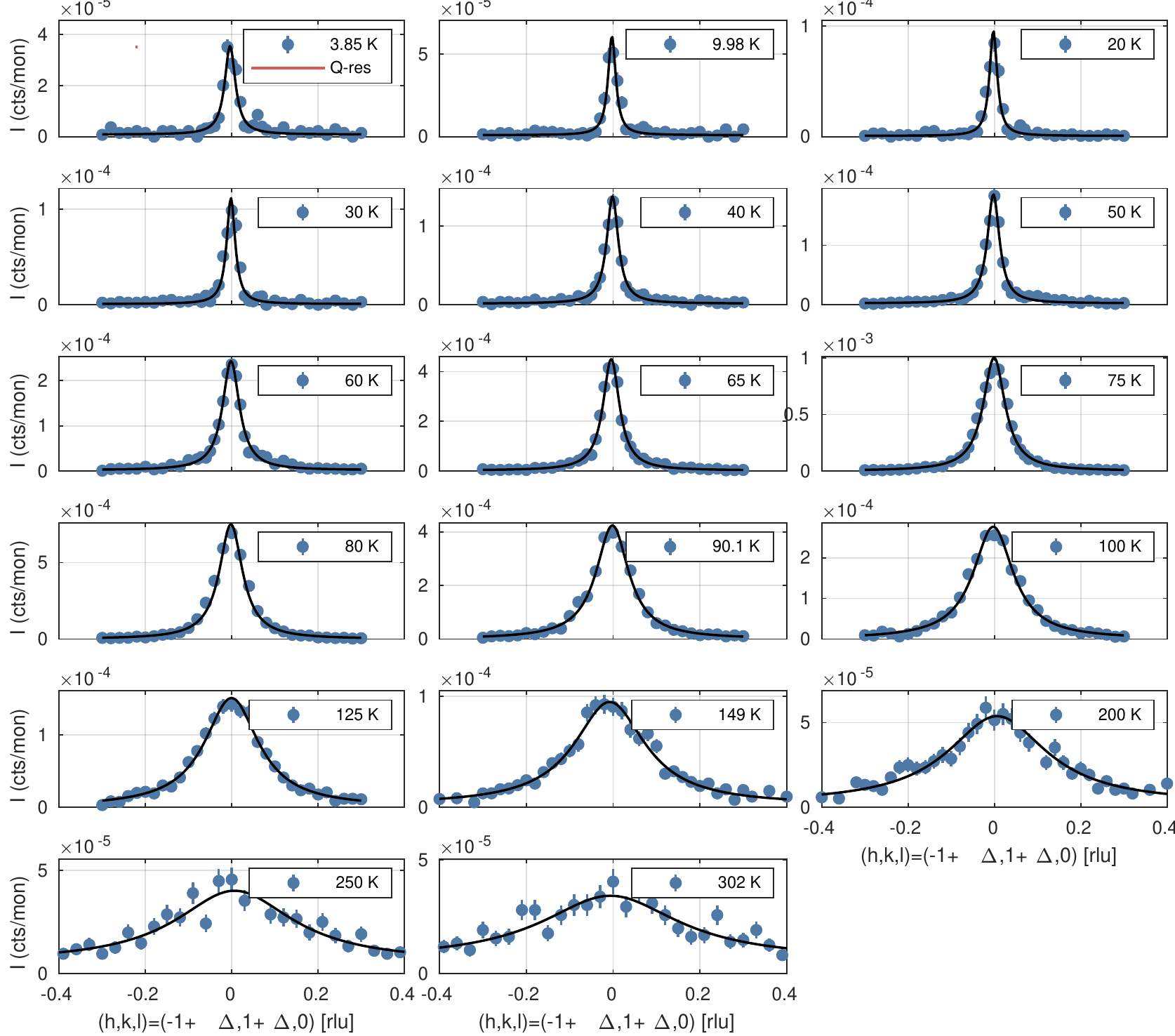}
    \caption{Raw scans for the $\Gamma^\prime$ point excitation at $\hbar \omega = 1$~meV taken with SPINS for various temperatures, along with Lorentzian fits.}
    \label{fig:Kscans}
\end{figure*}

For the diffuse excitations at the $\Gamma^\prime$ point a cold neutron TAS was needed with sufficient energy resolution to access the excitation in the ordered phase below the spin wave gap. Here we utilized the cold neutron TAS SPINS (NIST) with a closed-cycle cryostat. SPINS was configured with constant $E_f=5.0$~meV and a flat analyzer, as well as using a cold Be-filter.
To cross-normalize, a scan of the M-point at $T=100$~K was performed at SPINS, and the intensity was cross-normalized to the Eiger data. This normalization factor was then applied to all the scans made with SPINS, the results of which are shown in Figs. 2 and 4 in the main article. 
The raw scans and their Lorentzian fits can be seen in Fig. \ref{fig:Kscans}.

In order to investigate the dimensionality of the excitations, the $M^\prime$ excitation was measured for varying $l$. The measurement was performed on a different, smaller (1.3~g) sample, aligned in the $(hhl)$ scattering plane. The $M^\prime$ excitation was measured to avoid the complication of spin waves at $\Gamma^\prime$. The TAS measurements were made at IN3 (ILL). Scans were made across the $M^\prime$ point at $(hhl)=(\frac{1}{2} \frac{1}{2} l)$ with varying $l$ and a constant energy transfer of $\hbar\omega = 2$~meV. The width of the elastic peak varied considerably with $l$, so a background subtraction was performed using scans taken at the same $q$ positions at 2~K, where the M$^\prime$ signal is extinct. This data is shown in figure 4 in the main text.\\

%\FloatBarrier
\section{Simulation details}
The scattering is modeled as originating from spin clusters calculated via the static structure factor,

\begin{align}
       S(\mathbf{Q})= f(Q)^2 \sum_{l=0}^N  e^{i \mathbf{Q} \cdot \mathbf{r}_{l}}  \sum_{\substack{\alpha, \beta \\ \in \{x,y,z\}} } \left(\delta_{\alpha, \beta}-\hat{Q}_{\alpha} \hat{Q}_{\beta}\right)
 \left\langle S_{0}^{\alpha} S_{l}^{\beta}\right\rangle,
 \label{eq:SqApp}
\end{align}
where $N$ is the number of spins, $f(Q)$ is the magnetic form factor for Mn$^{3+}$, $\mathbf{r}_l$ their position in the lattice, and $\left\langle S_{0}^{\alpha} S_{l}^{\beta}\right\rangle$ represents a $4\pi$ rotational average of the spin-spin interaction. The spins are placed in the 120$^\circ$ ground state structure, and the rotational average represents a coherent rotation of all spins.

\begin{figure}[htb]
\subfloat{\includegraphics{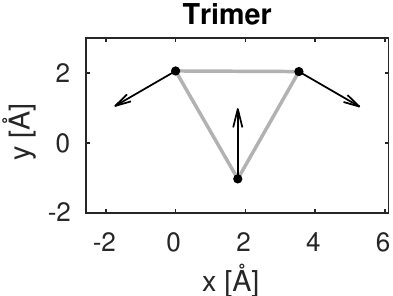}}
\quad
\subfloat{\includegraphics{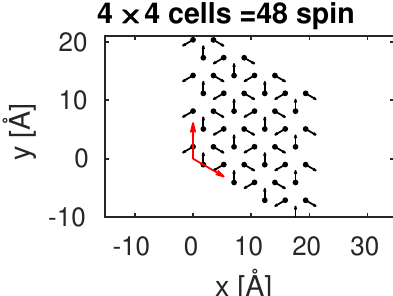}}
\caption{Spin position and spin direction in real space for the trimer spin cluster (left) and a 4$\times$4 unit cell cluster. Red arrows in the last subfigure indicate the unit vectors $\hat{\mathbf{a}}$ and $\hat{\mathbf{b}}$ in real-space.}
    \label{fig:pos}
\end{figure}

In the main text two examples of spin clusters are referred to: the minimal trimer model, which is illustrated in Fig. \ref{fig:pos}(a), and a much larger cluster with critical spin-spin correlations. An illustration of a $4\times 4$ unit cell cluster with 48 spins is shown in Fig. \ref{fig:pos}(b), although in the simulation a larger $40\times40$ unit cell is used. The critical spin-spin correlations are implemented by letting them exponentially decay, i.e. $\langle S_0 S_l \rangle \propto \exp(|\mathbf{r}_{0l}|/\xi)$, where $|\mathbf{r}_{0l}|$ is the distance between the 0'th and $l$th spin. 

\begin{figure}[htb]
\subfloat{\includegraphics{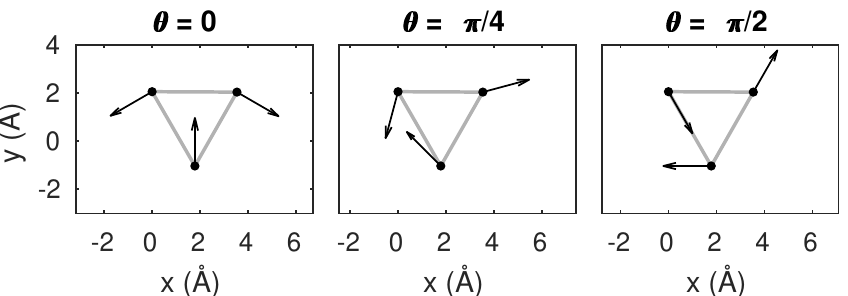}}

\subfloat{\includegraphics{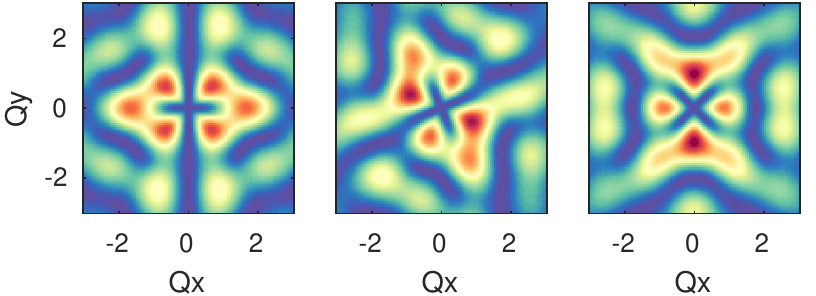}}
\caption{Calculation of the structure factor for the spin trimer. Upper row shows the instantaneous spin structure for three different rotations in the $xy$-plane, while lower row shows the associated snapshot of the structure factor, i.e. without the rotational average.}
    \label{fig:Rot}
\end{figure}

In order to make the steps in this simulation transparent, below we will highlight some details usually omitted from articles. Firstly, we wish to make clear that it is the rotational average that is responsible for obtaining the hexagonal symmetry of the structure factor. This illustrated in Fig. \ref{fig:Rot} for the simple trimer, where the upper row shows instantaneous spin configuration for different rotations in the $xy$-plane, and the lower row shows a snapshot of the calculated structure factor, i.e. eq.~(\ref{eq:SqApp}) without the rotational average. As can be seen, these snapshots do not possess the hexagonal symmetry of the underlying lattice, and it is the total rotational average that possesses the hexagonal symmetry.

\begin{figure}[ht!]
\subfloat{\includegraphics[width=0.49\linewidth]{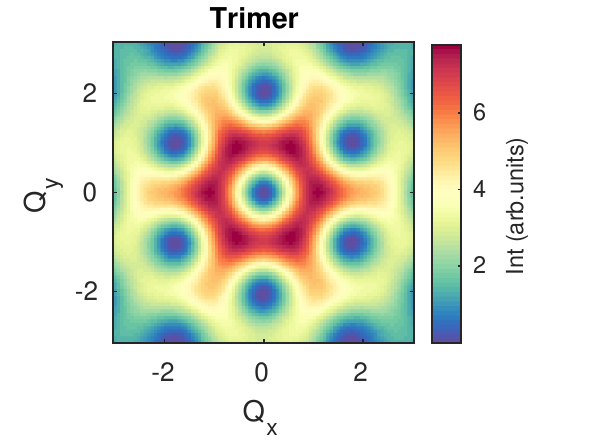}}
~
\subfloat{\includegraphics[width=0.49\linewidth]{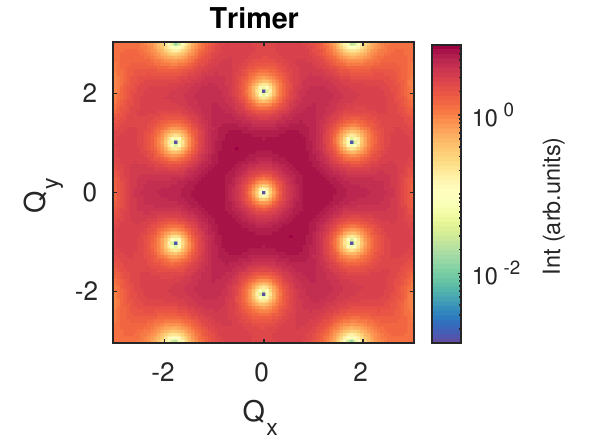}}

\subfloat{\includegraphics[width=0.49\linewidth]{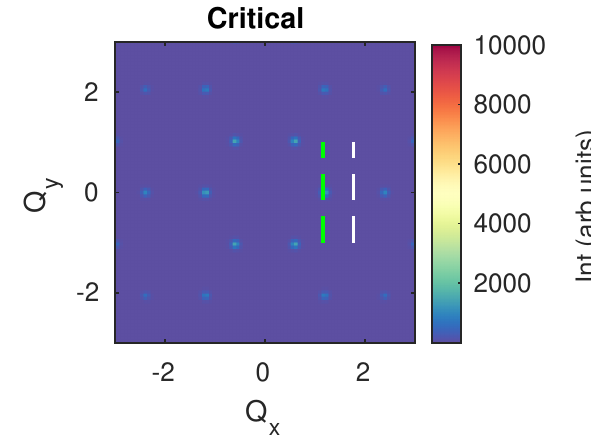}}
~
\subfloat{\includegraphics[width=0.49\linewidth]{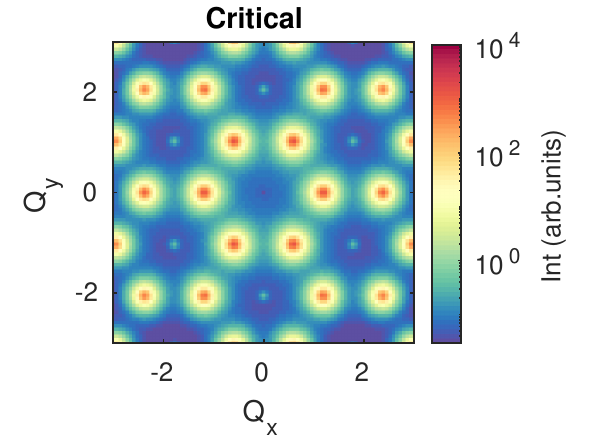}}
\caption{Structure factor for the trimer (top) and the $40 \times 40$ cluster with exponentially decaying spin-spin correlations with $\xi=8.6$~\AA~(bottom). Left side shows the intensity on a linear scale, while right side shows the intensity on logarithmic scale for both systems. Intensities are not scaled to match experimental data here.}
    \label{fig:FQ}
\end{figure}

For the trimer and the critical model, one can then calculate the cluster structure factor. For the critical model, $\xi=8.6$\AA~is used here. Both calculations are shown in Fig. \ref{fig:FQ} for a Q grid with $\Delta Q = 0.07$~\AA$^{-1}$~for both systems on both linear and logarithmic intensity scale for comparison. While the trimer structure factor on a linear scale may look somewhat like the experimental data for \ymo, it does not adequately capture large variations in intensity of more than 2-3 decades between the K and M$^\prime$ point, and is much more blob-like, whereas the model including critical fluctuations shown on the lower row reproduces experimental data more accurately on a logarithmic intensity scale.

\begin{figure}[htb]
\includegraphics{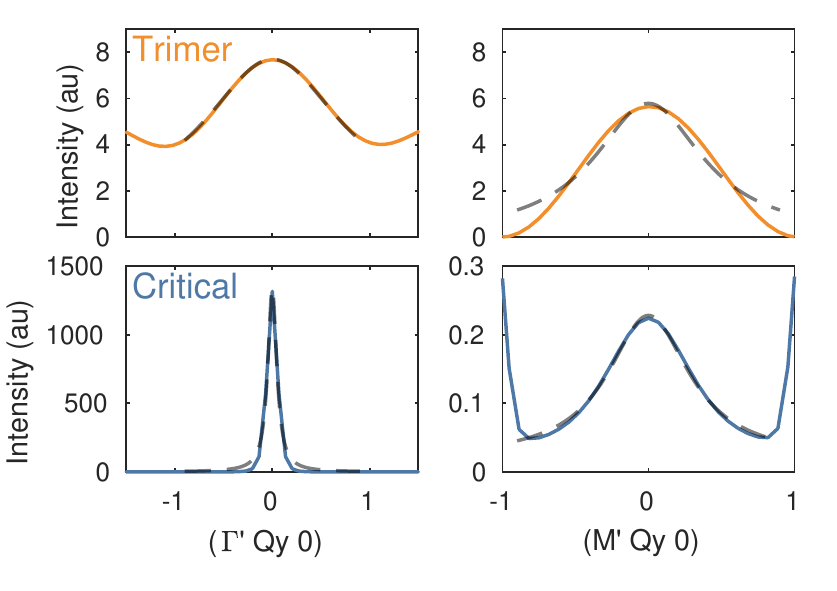}
\caption{One-dimensional cuts through the K and M$^\prime$ points for the trimer and $10\times10$ model. Upper row shows the trimer, while lower row shows data for the $10\times10$ model. Left hand side shows cuts through the $\Gamma^\prime$ point, while right hand side shows cuts through the M$^\prime$ point. The cutting directions are shown in Fig. \ref{fig:FQ} (lower left corner) as dashed lines. The black dashed lines superimposed on the data here are Lorentzian fits to data.}
    \label{fig:CutCompare}
\end{figure}

Another argument in favor of the critical model, as also mentioned in the main text, is that the intensity profiles are much better captured by the critical model than the trimer model. As this is hard to see on color maps,  Fig. \ref{fig:CutCompare} shows cuts through the $\Gamma^\prime$ and M$^\prime$ point for the two models. The cutting directions are indicated by the dashed lines in Fig. \ref{fig:FQ} (bottom left). The inner, green dashed line indicates a cut across the $\Gamma^\prime$ point, while the outer, white line indicates a cut across the M$^\prime$ point. Coming back to the data in Fig. \ref{fig:CutCompare}, the top row shows the trimer data, while the bottom row shows the critical model. All data has been fitted with a Lorentzian interval [-1,1]. The trimer model (top row) has very similar intensities at the K and M$^\prime$ points, and very broad peak profiles at both locations. This is opposed to the critical model (bottom row), where the peak shape is very different between the K and M$^\prime$ points: the $\Gamma^\prime$ point (left) is three orders of magnitudes more intense than the M$^\prime$ point, and narrower. This is a much better reproduction of the experimental data, which was shown in Figs. \ref{fig:Mscans} and \ref{fig:Kscans}. For this reason, we believe the critical model is a much closer reproduction of the real  system, and computationally not much more expensive.

%\bibliography{bib}

%\end{document}

\end{document}